\documentclass[twocolumn]{aastex631} 
\usepackage[T1]{fontenc}
\usepackage[utf8]{inputenc}
\usepackage{cancel}

\usepackage{verbatim}
\usepackage{needspace}

\definecolor{my_color}{HTML}{3a18b1}

\definecolor{new_color}{HTML}{CF0000}

\definecolor{html_black}{HTML}{000000}

\definecolor{new_color}{HTML}{000000}

\newcommand\redit[1]{\textcolor{new_color}{#1}}

\shorttitle{VOYAGERS Paper 1: Survey Design}
\shortauthors{Aloisi et al.}
\graphicspath{{./}{figures/}}

\begin{document}

\title{Searching for Exoplanets Born Outside the Milky Way: VOYAGERS Survey Design}

\correspondingauthor{Robert Aloisi}
\email{rjaloisi@wisc.edu}

\author[0000-0003-2822-616X]{Robert Aloisi}
\affiliation{Department of Astronomy,  University of Wisconsin-Madison, 475 N.~Charter St., Madison, WI 53706, USA}

\author[0000-0001-7246-5438]{Andrew Vanderburg}
\affiliation{Center for Astrophysics ${\rm \mid}$ Harvard {\rm \&} Smithsonian, 60 Garden Street, Cambridge, MA 02138, USA}
\affiliation{Department of Physics \& Kavli Institute for Astrophysics and Space Research, Massachusetts Institute of Technology, Cambridge, MA 02139, USA}

\author[0000-0001-7493-7419]{Melinda Soares-Furtado}
\affiliation{Department of Astronomy,  University of Wisconsin-Madison, 475 N.~Charter St., Madison, WI 53706, USA}
\affiliation{Department of Physics, University of Wisconsin--Madison, 1150 University Ave, Madison, WI 53706, USA}

\author[0000-0002-1617-8917]{Phillip Cargile}
\affiliation{Center for Astrophysics ${\rm \mid}$ Harvard {\rm \&} Smithsonian, 60 Garden Street, Cambridge, MA 02138, USA}

\author[0000-0002-0661-7517]{Ke Zhang}
\affiliation{Department of Astronomy,  University of Wisconsin-Madison, 475 N.~Charter St., Madison, WI 53706, USA}

\author[0000-0003-2806-1414]{Lina Necib}
\affiliation{Department of Physics \& Kavli Institute for Astrophysics and Space Research, Massachusetts Institute of Technology, Cambridge, MA 02139, USA}

\author[0000-0001-9911-7388]{David W. Latham}
\affiliation{Center for Astrophysics ${\rm \mid}$ Harvard {\rm \&} Smithsonian, 60 Garden Street, Cambridge, MA 02138, USA}

\author[0000-0002-8964-8377]{Sam Quinn}
\affiliation{Center for Astrophysics ${\rm \mid}$ Harvard {\rm \&} Smithsonian, 60 Garden Street, Cambridge, MA 02138, USA}

\author[0000-0002-1533-9029]{Emily Pass}
\affiliation{Kavli Institute for Astrophysics and Space Research, Massachusetts Institute of Technology, Cambridge, MA 02139, USA}

\author[0000-0002-1092-2995]{Anne Dattilo}
\affiliation{Center for Exoplanets and Habitable Worlds, Penn State University, 525 Davey Laboratory, 251 Pollock Road, University Park, PA, 16802, USA}
\affiliation{Department of Astronomy \& Astrophysics, University of California Santa Cruz, 1156 High Street, Santa Cruz, CA, 95064, USA}
\email{adattilo@psu.edu}

\author[0000-0002-6871-6131]{Giacomo Mantovan}
\affiliation{Centro di Ateneo di Studi e Attività Spaziali ``G. Colombo'' -- Università degli Studi di Padova, Via Venezia 15, IT-35131, Padova, Italy}

\author[0000-0003-1316-1033]{Francesco Amadori}
\affiliation{INAF - Osservatorio Astrofisico di Torino, via Osservatorio 20,10025 Pino Torinese, Italy}
\affiliation{Dipartimento di Fisica e Astronomia ``Galileo Galilei'', Università di Padova, Vicolo dell'Osservatorio 3, IT-35122, Padova, Italy}

\author[0000-0003-4903-567X]{Mariona Badenas-Agusti}
\affiliation{Institute of Astronomy, University of Cambridge, Madingley Road, Cambridge, CB3 0HA, UK}
\affiliation{Institut d'Estudis Espacials de Catalunya, 08860 Castelldefels, Barcelona, Spain}

\author[0009-0005-7108-9502]{Perry Berlind}
\affiliation{Center for Astrophysics ${\rm \mid}$ Harvard {\rm \&} Smithsonian, 60 Garden Street, Cambridge, MA 02138, USA}

\author[0000-0003-4830-0590]{Francesco Borsa}
\affiliation{INAF -- Osservatorio Astronomico di Brera, Via E. Bianchi 46, 23807 Merate (LC), Italy}

\author[0000-0001-9978-9109]{Walter Boschin}
\affiliation{Fundacion Galileo Galilei - INAF (Telescopio Nazionale Galileo): Breña Baja, ES}

\author[0000-0002-5130-4827]{Lorenzo Cabona}
\affiliation{INAF -- Osservatorio Astronomico di Brera, Via E. Bianchi 46, 23807 Merate (LC), Italy}

\author[0000-0002-2830-5661]{Michael L. Calkins}
\affiliation{Center for Astrophysics ${\rm \mid}$ Harvard {\rm \&} Smithsonian, 60 Garden Street, Cambridge, MA 02138, USA}

\author[0000-0003-0047-4241]{Hans J. Deeg}
\affiliation{\label{ins:iac} Instituto de Astrof\'isica de Canarias (IAC), 38205 La Laguna, Tenerife, Spain}
\affiliation{\label{ins:ull} Departamento de Astrof\'isica, Universidad de La Laguna (ULL), 38206 La Laguna, Tenerife, Spain}

\author[0000-0002-9332-2011]{Xavier Dumusque}
\affiliation{Astronomy Department of the University of Geneva, 51 chemin des Maillettes, CH-1290 Versoix, Switzerland}

\author[0000-0002-9789-5474]{Gilbert A. Esquerdo}
\affiliation{Center for Astrophysics ${\rm \mid}$ Harvard {\rm \&} Smithsonian, 60 Garden Street, Cambridge, MA 02138, USA}

\author[0009-0006-6397-2503]{Yoshi Nike Emilia Eschen}
\affiliation{Department of Physics, University of Warwick, Gibbet Hill Road, Coventry CV4 7AL, United Kingdom}

\author[0009-0000-5623-5237]{S.~Filomeno}
\affiliation{INAF-Osservatorio Astronomico di Roma, Via Frascati 33, I-00040 Monte Porzio Catone (RM), Italy Dipartimento di Fisica}
\affiliation{Universit\`a di Roma Tor Vergata, Via della Ricerca Scientifica 1, I-00133 Roma, Italy}
\affiliation{Dipartimento di Fisica, Sapienza Università di Roma, Piazzale Aldo Moro 5, I-00185 Roma, Italy}

\author[0009-0002-5545-3034]{S. Gerald\'ia-Gonz\'alez}
\affiliation{\label{ins:iac} Instituto de Astrof\'isica de Canarias (IAC), 38205 La Laguna, Tenerife, Spain}
\affiliation{\label{ins:ull} Departamento de Astrof\'isica, Universidad de La Laguna (ULL), 38206 La Laguna, Tenerife, Spain}

\author[0000-0002-5169-9427]{Natalia Guerrero}
\affiliation{Department of Astronomy, University of Florida, Bryant Space Science Center, Stadium Road, Gainesville, FL 32611, USA}

\author[0000-0001-9827-1463]{Sydney Jenkins}
\affiliation{Department of Physics \& Kavli Institute for Astrophysics and Space Research, Massachusetts Institute of Technology, Cambridge, MA 02139, USA}

\author[0000-0002-4197-7374]{Gaia Lacedelli}
\affiliation{\label{ins:iac} Instituto de Astrof\'isica de Canarias (IAC), 38205 La Laguna, Tenerife, Spain}

\author[0000-0003-3204-8183]{Mercedes L\'opez-Morales}
\affiliation{Space Telescope Science Institute, 3700 San Martin Drive, Baltimore MD 21218, USA}

\author[0009-0001-3400-6705]{Paula Manuela Leguizamon-Pineda}
\affiliation{Dipartimento di Fisica e Astronomia ``Galileo Galilei'', Università di Padova, Vicolo dell'Osservatorio 3, IT-35122, Padova, Italy}

\author[0000-0002-1742-7735]{Emilio Molinari}
\affiliation{INAF – Osservatorio Astronomico di Cagliari, via della Scienza 5, Selargius (CA), Italy}


\author[0000-0003-0061-518X]{Juan Carlos Morales}
\affiliation{Institut d'Estudis Espacials de Catalunya, 08860 Castelldefels, Barcelona, Spain}
\affiliation{Institut de Ci\`encies de l'Espai (CSIC), de Can Magrans, Campus UAB, 08193 Bellaterra, Barcelona, Spain}

\author[0000-0001-7254-4363]{Annelies Mortier}
\affiliation{School of Physics \& Astronomy, University of Birmingham, Edgbaston, Birmingham, B15 2TT, UK}

\author[0000-0003-3997-5705]{Rohan Naidu}
\affiliation{Department of Physics \& Kavli Institute for Astrophysics and Space Research, Massachusetts Institute of Technology, Cambridge, MA 02139, USA}

\author[0000-0003-1149-3659]{Domenico Nardiello}
\affiliation{Dipartimento di Fisica e Astronomia ``Galileo Galilei'', Università di Padova, Vicolo dell'Osservatorio 3, IT-35122, Padova, Italy}

\author[0000-0003-1360-4404]{Belinda Nicholson}
\affiliation{Centre for Astrophysics, University of Southern Queensland, West St., Toowoomba, Queensland, Australia, 4350}

\author[0000-0001-9573-4928]{Isabella Pagano}
\affiliation{INAF, Osservatorio Astrofisico di Catania, Via S. Sofia 78, 95123 Catania, Italy}

\author[0000-0003-0987-1593]{Enric Palle}
\affiliation{\label{ins:iac} Instituto de Astrof\'isica de Canarias (IAC), 38205 La Laguna, Tenerife, Spain}
\affiliation{\label{ins:ull} Departamento de Astrof\'isica, Universidad de La Laguna (ULL), 38206 La Laguna, Tenerife, Spain}

\author[0000-0002-5752-6260]{Marco Pedani}
\affiliation{Fundación Galileo Galilei - INAF: Breña Baja, Santa Cruz de Tenerife, ES}

\author[0000-0002-4445-1845]{Matteo Pinamonti}
\affiliation{INAF - Osservatorio Astrofisico di Torino, via Osservatorio 20,10025 Pino Torinese, Italy}

\author[0000-0002-2218-5689]{Jesus Maldonado}
\affiliation{INAF - Osservatorio Astronomico di Palermo: Palermo, IT}

\author[0000-0002-8786-2572]{Monica Rainer}
\affiliation{INAF Osservatorio Astronomico di Brera
Via E. Bianchi, 46, 23807 Merate (LC), Italy}

\author[0000-0002-6689-0312]{Ignasi Ribas}
\affiliation{Institut d'Estudis Espacials de Catalunya, 08860 Castelldefels, Barcelona, Spain}
\affiliation{Institut de Ci\`encies de l'Espai (CSIC), de Can Magrans, Campus UAB, 08193 Bellaterra, Barcelona, Spain}

\author[0000-0002-6379-9185]{Ken Rice}
\affiliation{Institute for Astronomy, University of Edinburgh, Royal Observatory, Blackford Hill, Edinburgh, EH9 3HJ, UK}
\affiliation{Centre for Exoplanet Science, University of Edinburgh, Edinburgh, EH9 3HJ, UK}

\author[0000-0001-7409-5688]{Guðmundur Kári Stefánsson}
\affiliation{Anton Pannekoek Institute for Astronomy, University of Amsterdam, Science Park 904, 1098 XH Amsterdam, The Netherlands}

\author[0009-0001-0867-9711]{Daisy Turner}
\affiliation{School of Physics \& Astronomy, University of Birmingham, Edgbaston, Birmingham, B15 2TT, UK}

\author[0000-0001-8749-1962]{Thomas G. Wilson}
\affiliation{Department of Physics, University of Warwick, Gibbet Hill Road, Coventry CV4 7AL, United Kingdom}

\author[0000-0002-6532-4378]{Mathias Zechmeister}
\affiliation{Institut f\"ur Astrophysik und Geophysik, Georg-August-Universit\"at, Friedrich-Hund-Platz 1, 37077 G\"ottingen, Germany}

\begin{abstract}
Observations over the past few decades have found that planets are common around nearby stars in our Galaxy, but little is known about planets that formed outside the Milky Way. We describe the design and early implementation of a survey to test whether planets also exist orbiting the remnant stars of ancient dwarf galaxies that merged with the Milky Way, and if so, how they differ from their Milky Way counterparts. 
VOYAGERS (\textit{Views Of Yore - Ancient Gaia-enceladus Exoplanet Revealing Survey}) is a radial velocity (RV) search using precision spectrographs to discover exoplanets orbiting very low metallicity ($-2.8 < [\mathrm{Fe/H}] \leq -0.8$) stars born in the dwarf galaxy Enceladus, which merged with the Milky Way galaxy about 10 Gyr ago. A sample of 22 candidates have been screened from a catalog of Gaia-Enceladus-Sausage (GES) members using a combination of stellar properties and reconnaissance observations from the TRES spectrograph. Precision RV measurements have been initiated using the NEID, HARPS-N, and CARMENES spectrographs.  We plan to focus most upcoming observations on 10 main sequence targets.  Data collection is well underway, with 778 observations on 22 candidates (385 of which are 10 focus targets), but far from complete.  This survey is designed to be sensitive to sub-Neptune mass planets with periods up to hundreds of days.  We note that the RV analysis gives mass multiplied by $\sin (inclination)$ or the minimum mass for exoplanets.  The expected survey yield is three planets, assuming that occurrence rates are similar to those in the Milky Way and taking into account the degeneracy with inclination in our yield models.  Our survey is designed to detect at least one exoplanet if occurrence rates are similar to known Milky Way exoplanets or, if no exoplanets are discovered, to rule out a Milky Way-like planet population in GES with 95\% confidence level.   

\end{abstract}

\keywords{Exoplanet detection (490), Radial Velocity (1332), Dwarf Galaxies (416)}


\section{Introduction}

More than 6000 confirmed exoplanets have now been found orbiting stars in the Milky Way, with projections that most Milky Way stars host planets \citep{2013ApJ...766...81F}.  
The two most common exoplanet detection techniques are radial velocity \redit{(RV)} surveys and transit searches. \redit{RV} surveys---such as the California Legacy Survey \citep{2021ApJS..255....8R} and \redit{the HARPS survey} \citep{2011arXiv1109.2497M}---measure Doppler shifts in stellar spectral lines to infer the Keplerian motion induced by orbiting planets, while transit searches---exemplified by Kepler \citep{2011ApJ...736...19B} and TESS \citep{2015JATIS...1a4003R}---detect the slight dip in stellar brightness when a planet transits its host star.
Most known exoplanets have been discovered orbiting main sequence stars in the Milky Way disk, which typically have metallicities similar to that of the Sun. 
However, this census of known exoplanet hosts does not fully represent the wider diversity of stars across the Universe, including metal-poor stars from the early Universe and stars found in dwarf galaxies.  This may in part be due to observational bias for \redit{RV} surveys because metal-poor stars may be eliminated because they have fewer spectral lines and thus lower precision compared to metal-rich stars.  An exception to this bias is a HARPS survey targeting 109 metal-poor stars, which found an exoplanet orbiting a thin disk star with [Fe/H] of $-0.62$ \citep{Mortier2016A&A}.

We speculate that some planets likely formed in the low-metallicity, high-alpha element environment (elements formed by fusion of He nuclei) of the early Universe, and this population may differ in occurrence rates and compositions compared to those found in more recently formed stars in the Milky Way disk. For example, occurrence rates for exoplanets with masses larger than Jupiter tend to be lower around low-metallicity stars \citep{2005ApJ...622.1102F, 2018AJ....155...89P}. In contrast, occurrence rates for Neptune-sized and smaller planets show little dependence on metallicity \citep{2011A&A...526A.112S,2019Geosc...9..105A}.  Regarding exoplanet properties, the density of sub-Neptune planets has been found to be lower for low-metallicity stars \citep{2022MNRAS.511.1043W}.
Short-period super-Earths also appear to be relatively rare around metal-poor stars. A recent study constrained their occurrence rate to less than $1.67\%$ for stars with $-0.75 < \mathrm{[Fe/H]} \leq -0.50$, based on no detections in a TESS sample \citep{2024AJ....168..128B}. However, other surveys have identified super-Earths in this metallicity regime, such as K2-344\,b \citep{deLeon2021MNRAS}, HD 175607\,b \citep{Mortier2016A&A}, and HD 39194\,b/c/d \citep{Unger2021A&A}.

Higher alpha-element abundances may also influence the planet population. For example, alpha-enhanced stars have been found to host small planets at significantly higher rates. \citet{2012A&A...547A..36A} reported an occurrence rate of 12\% for small planets around low-metallicity ($-0.65 < \mathrm{[Fe/H]} \leq-0.30$) and high-alpha ($\mathrm{[Ti/Fe]} > 0.2$) stars, compared to just 2\% for low metallicty planet-hosting stars that are not alpha-enhanced. Similarly, \citet{2018ApJ...867L...3B} found that compact multi-planet systems might occur three times more frequently around stars with [Fe/H] ${\sim}-0.5$. \redit{Theoretical estimates of exoplanet occurrence rates (total number of this type planets divided by total number of target stars) around low-metallicity halo stars from \citet{2024A&A...692A.150B} (based on the New Generation Planet Population Synthesis code, \citealt{Emsenhuber2021A&A}) are 0.8\% for Super-Earths, 6.0\% for Earths, and 0.0\% Giant planets.}  These findings suggest that the early Universe’s low-metallicity, high-alpha environment may have produced a distinctly different planet population.

Other studies suggest that metal-poor stars---despite having fewer planet-forming materials---may produce more compact multi-planet systems. Observations indicate that these systems often consist of planets of similar mass arranged in regular intervals, a phenomenon sometimes described as ``peas in a pod'' \citep{2018AJ....155...48W}. Supporting this pattern, theoretical work by \citet{2020MNRAS.493.5520A} shows that when the total planet mass is below $40\,\mathrm{M_{\oplus}}$, adjacent planets are likely to form with comparable masses and evenly spaced orbits. In contrast, systems with total masses exceeding $40$\,M$_{\oplus}$ tend to concentrate most of the mass in a single dominant planet.

However, these studies have focused on stars that formed within the Milky Way disk and \redit{that} have metallicities not so different from the Sun.  New opportunities have arisen to study planets forming in more extreme metallicity and alpha-abundance environments. Thanks to ESA’s \textit{Gaia} mission \citep{GaiaCollaboration2016A&A}, several studies have identified stellar streams that likely originated outside the Milky Way. These features were distinguished by their distinct kinematics and lower metallicities compared to Milky Way disk stars. One notable example is Gaia-Enceladus-Sausage (GES), a structure produced by the merger of a ${\sim}10^8\,\mathrm{M_{\odot}}$ dwarf galaxy with the Milky Way about 10 billion years ago \citep{2018MNRAS.478..611B,2018Natur.563...85H}. The stars from GES formed in a Universe with lower overall metallicity, resulting in reduced [Fe/H] abundances, yet they remain relatively enriched in alpha elements.
This alpha enhancement arises because alpha elements are predominantly produced in Type II supernovae, which occur earlier in a galaxy’s evolution compared to Type Ia supernovae. Moreover, GES stars are widely distributed throughout the Milky Way, with a substantial number of members located near the Sun. Many of these stars are sufficiently bright that an \redit{RV} survey could feasibly detect sub-Neptune mass planets \citep{2020ApJ...903...25N}.

Searching for planets in GES presents an intriguing opportunity, as it remains unclear how planets form in environments outside the Milky Way and how low-metallicity conditions influence these processes. A recent search \redit{of data from the Transiting Exoplanet Survey Satellite (TESS)} for transiting planets within GES \citep{2022AJ....164..119Y} found no detections, but the sensitivity \redit{of this search} was limited to giant planets with periods under 10 days, and its sample size of 1,080 stars was relatively small for a transit survey. Despite these null results, evidence from similar environments, such as thick-disk stars, suggests that planet formation can occur under low-metallicity conditions. 

These findings highlight the potential for future observations to shed light on planet formation in such extreme environments.

To address these open scientific questions, we have initiated the VOYAGERS survey (\textit{Views Of Yore --- Ancient Gaia-enceladus Exoplanet Revealing Survey}). 
Using the RV technique, our survey focuses on studying main sequence and slightly evolved stars in the Gaia-Enceladus dwarf galaxy. Our primary goal is to detect exoplanets that formed in low-metallicity environments beyond the Milky Way. 

The structure of this paper is as follows. In Section~\ref{section:targets}, we outline the target selection process and describe the stellar properties of our sample. Section~\ref{section:yield} presents survey simulations and discusses the optimizations used to maximize detection yield. Section~\ref{section:rv} details the RV observations, and Section~\ref{section:search} explains the search methodology. In Section~\ref{section:discussion}, we describe injection / recovery  tests to evaluate our survey’s sensitivity and assess the progress made. 
We summarize our findings and conclusions in Section~\ref{section:conclusions}.

\section{Target Selection}
\label{section:targets}

\subsection{Gaia-Enceladus Catalog} \label{GEScatalog}

\begin{figure*}[!tbh]
\centering{\includegraphics[width=\textwidth]{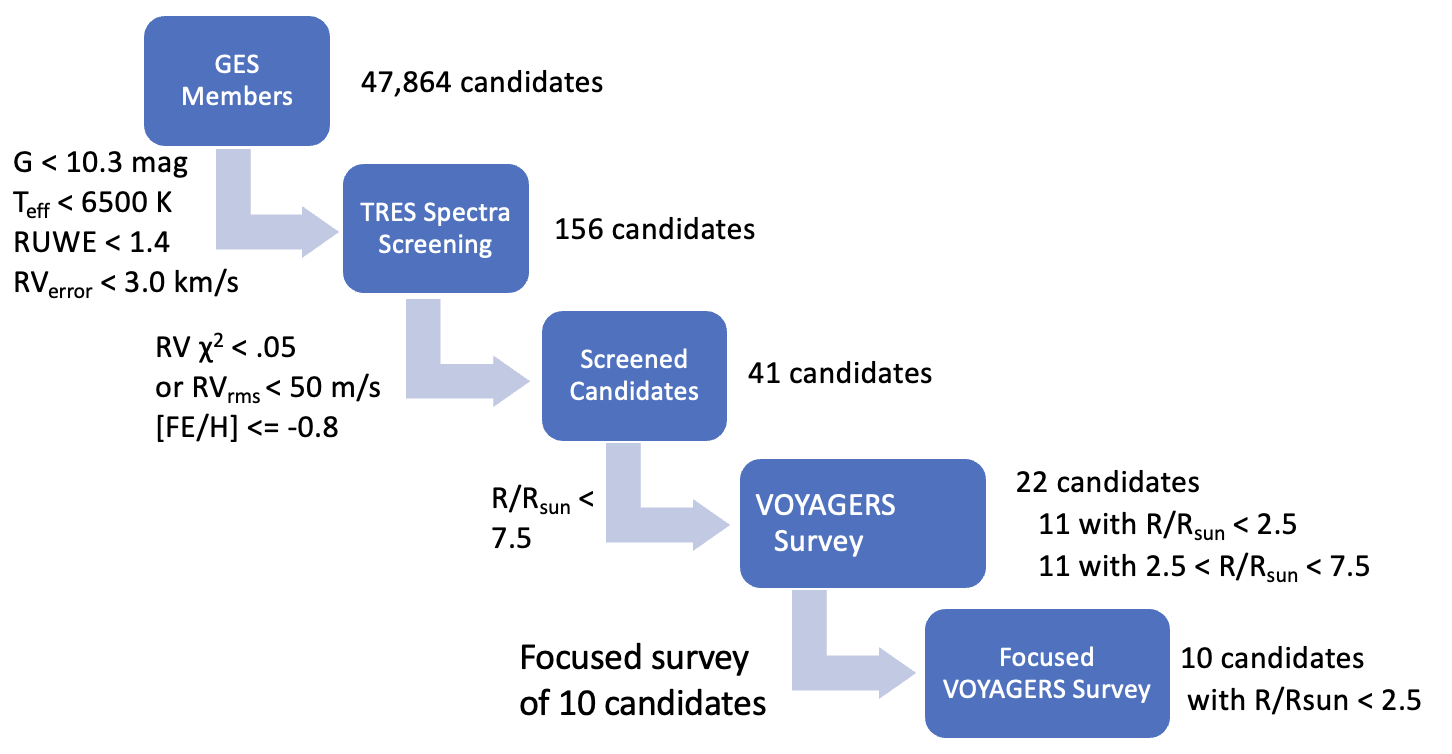}}
\caption{VOYAGERS target stars were selected from 47,864 GES members following several screening steps. Available stellar properties were used to screen to 156 candidates.  TRES spectra were then used to identify 41 candidates that would be best for a precision RV survey.  Our initial survey includes 11 high priority main sequence ($R/\mathrm{R_\odot}<2.5$) and 11 more evolved ($R/\mathrm{R_\odot}< 7.5$) targets.  We plan to focus more observations on 10 main sequence targets for upcoming observations.\\}
\label{fig:selection}
\end{figure*}

Figure~\ref{fig:selection} provides a summary of our candidate selection process. Our target list includes stars identified as members of GES, with membership determined using the method outlined in \citet{2020ApJ...903...25N}. This approach incorporated a neural network to classify accreted stars in \citet{2020A&A...636A..75O} based on \textit{Gaia} DR2 data \citep{2018A&A...616A...1G} for which RV measurements were available.  
This paper used clustering algorithms to identify structures such as GES.  

Starting with the 47,864 stars whose GES membership probability exceeded 95\%, we then applied the following screening steps to define the VOYAGERS sample:
\begin{itemize}
    \item $Gaia\,G < 10.3$ (bright enough to ensure reasonable RV exposure times)
    \item $Gaia\,G < (Gaia (Bp - Rp) * 6.154 - 3.85$) (cut across color magnitude diagram to screen out more evolved Giant stars
    \item \redit{$Gaia\,T_{\mathrm{eff}} < 6500\,K$} (limited population to FGK-type stars, which are most suitable for a precision RV survey)
    \item $-15^{\circ} < Dec < 75^{\circ}$ (Declination suitable for Northern telescope survey)
    \item Gaia Renormalized Unit Weight Error (RUWE) $<$ 1.4 and Gaia \redit{$RV_{\mathrm{error}}< 3.0\,\mathrm{km\,s^{-1}}$} (to screen out binaries)

\end{itemize}
These selection criteria yielded a sample of 156 stars to include in our reconnaissance observations.

\subsection{Reconnaissance Observations}

Working from the list of candidates outlined in Section~\ref{GEScatalog}, we obtained reconnaissance spectra of 156 stars with the Tillinghast Reflector Echelle Spectrograph (TRES) on the 1.5-m telescope at Fred L.~Whipple Observatory (FLWO). These observations screened our targets for evidence of binarity, excessive stellar jitter, and other noise sources that could impede exoplanet detection.

TRES achieves an average spectral resolution of 44,000 and covers a wavelength range of $370-900$\,nm. 
Further, with exposures ranging between $600-1800\,s$, the instrument achieves a precision of approximately 30\,m\,s$^{-1}$ for the stars in our sample.
For each star, we secured at least four multi-order RV measurements over about a 6-month period and applied the selection procedure of Section~4.1 of \citet{2023AJ....166...11P}, which requires no significant variability (P(${\chi}^2$) $>$ 0.05).  We also cut targets with the overall root mean square velocity scatter less than 50\,m\,s$^{-1}$. We included the latter criterion because if we detected significant variability at such a low amplitude, it could indicate the presence of orbiting planets.  We identified a total of 41 stars that met these criteria.  Two stars, TIC 4624167 and TIC 72052060, had available archival HIRES RV data, so we did not conduct additional TRES observations on those stars.

\subsection{Stellar Properties}

We determined stellar parameters with a two-step process using the TRES spectra. We started by analyzing the spectra using the Stellar Parameters Classification (SPC) code \citep{2015ApJ...808..187B} to estimate the spectroscopic properties of our targets.  SPC compares spectra in the wavelength range between $505-536$\,nm to synthetic spectra spanning a grid of stellar parameters \citep{1992IAUS..149..225K} (including effective temperature, metallicity, surface gravity, and rotational velocity). The grid spans $3500 \, \mathrm{K} < T_{\text{eff}} < 9750 \, \mathrm{K}$, $0.0 < \log(g) < 5.0$, $-2.5 < \mathrm{[m/H]} < +0.5$ and rotational velocity of $0 \, \mathrm{km \, s^{-1}} < V_{\text{rot}} < 200 \, \mathrm{km \, s^{-1}}$ and has a spacing of $250 \, \mathrm{K}$ in effective temperature, $0.5$ in $\log(g)$, $0.5 \, \mathrm{dex}$ in $\mathrm{[m/H]}$ and progressive spacing in rotational velocity.  The peak height of the spectral cross correlation function (CCF) as a function of these different parameters is interpolated using polynomial models of the peak height as well as incorporating Yonsei-Yale (YY) isochrones \citep{2004ApJS..155..667D} to identify the best stellar parameters. 

While SPC was able to give initial estimates of the parameters, it is not optimal for stars in GES because it assumes an elemental abundance pattern identical to the Sun, while GES stars tend to be enhanced in alpha elements. We therefore performed analysis using \texttt{MINESweeper} \citep{2020ApJ...900...28C} stellar classification software incorporating TRES spectra, Gaia photometry and parallax, as well as MIST isochrones \citep{2016ApJS..222....8D,2016ApJ...823..102C,2015ApJS..220...15P}, which allows it to provide physical properties, such as stellar mass and radius, in addition to spectroscopic properties like [Fe/H], [$\alpha$/Fe] and lower $\log g$ values for evolved stars. \texttt{MINESweeper} uses a neural network trained with a parameter matrix including alpha metallicities up to $\mathrm{+0.6}$,and spectral resolutions up to 65,000. \texttt{MINESweeper} allows us to further distinguish thick and thin disk stars from GES stars. We ran \texttt{MINESweeper} using the SPC parameters as starting guesses for the fits.  We also used the prior parameters outlined in Table \ref{deluxetable:UberMS_priors}. These priors are based on those used by \citet{2025arXiv250618961P}.   We performed an initial fit with the standard \texttt{MINESweeper} package and then used the faster \texttt{uberMS}
\footnote{\href{hhttps://github.com/pacargile/uberMS}
{https://github.com/pacargile/uberMS}} version of the software to conduct a full fit (Cargile et al. \textit{in prep}).  The input spectra for this iteration were co-added from all available TRES observations, including two orders bridging the wavelength range $510-535\,nm$.

\begin{deluxetable*}{lc}
\label{deluxetable:UberMS_priors}
\tablecaption{Priors Used for \texttt{UberMS} Analysis\label{tab:parameters}}
\tablewidth{0pt}

\tablehead{
\colhead{Parameter} & \colhead{Prior}}
\startdata
Initial Mass, $M$/M$_\odot$ & IMF(0.60,0.95) \\
Initial [Fe/H] & $\mathcal{U}(\mathrm{-4.0,0.5})$ \\
Initial [$\alpha$/Fe] & U(-0.2,0.7) \\
Equivalent Evolutionary Point (EEP) & TN(250,50,200, 600) \\
log(Age),Gyr & double sigmoid(200.0,10.0,-200.0,10.05,9.25,11.25) \\
Stellar Rotation Velocity, km s$^{-1}$ & U(0,5) \\
A$_{V}$ & TN(dust map est.,err est.,0.0,0.5) \\
Distance, pc & TN(parallax est.,1,0,500) \\
\enddata
\tablecomments{IMF is the Kroupa Initial Mass Function \citep{2002Sci...295...82K}}

\tablecomments{Prior parameters used for \texttt{UberMS} stellar property fits.  For the priors, TN($\mu,\sigma$;a,b) is a truncated normal distribution with lower and upper limits given by a and b; U(a,b) stands for a uniform distribution between a and b; double sigmoid with (left slope, left break, right slope, right break, lower limit, upper limit).  Where indicated in the table, an estimate from previous fits was used in a truncated normal distribution}.
\end{deluxetable*}

\begin{figure*}[!ht]
\centering{\includegraphics[width=0.97\textwidth]{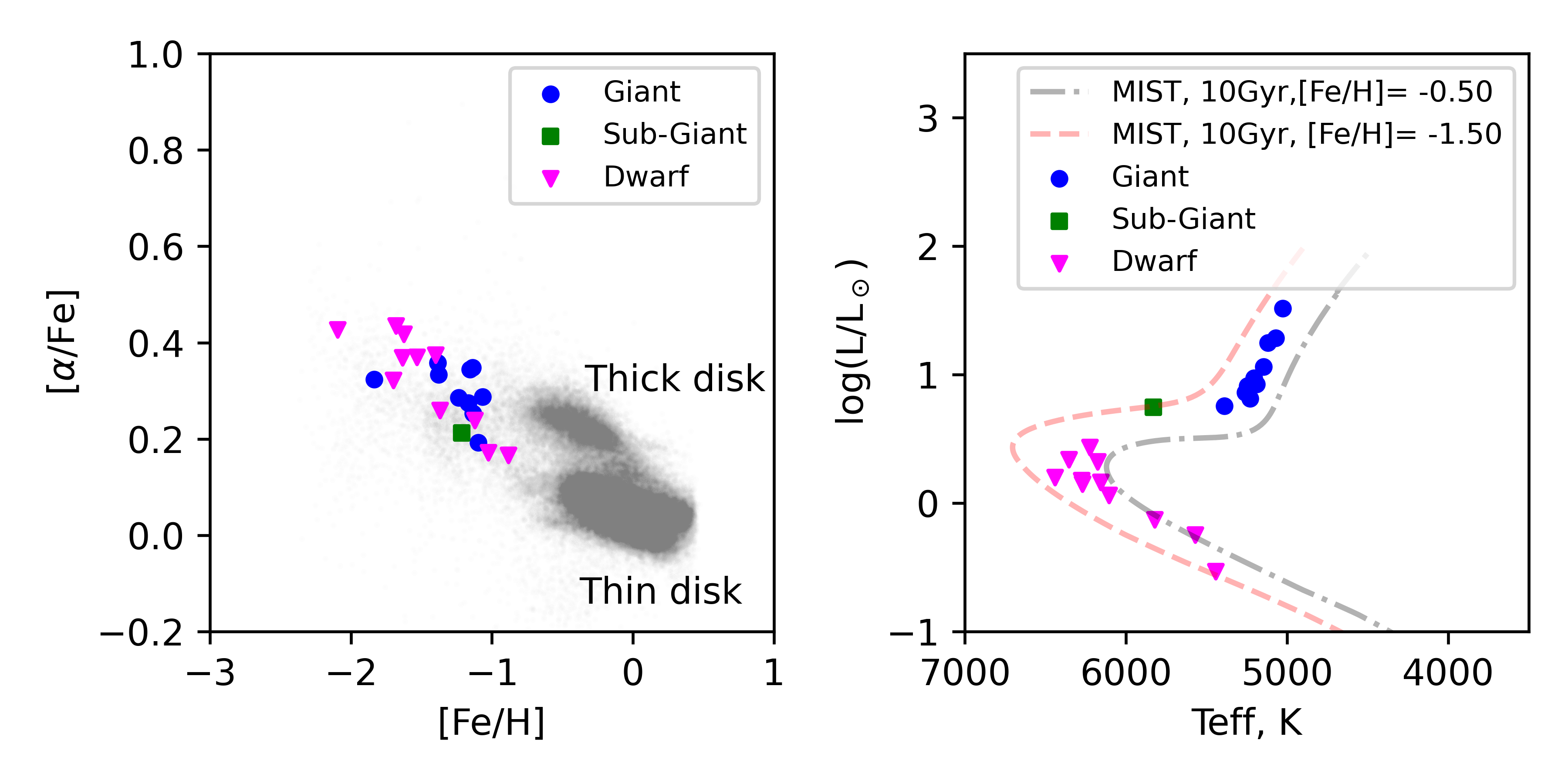}}
\caption{\textit{Left:} The metallicity properties of our survey stars are compared to typical properties of MW thin and thick disk stars.  The grey points are MW stars from the Apache Point Observatory Galactic Evolution Experiment (APOGEE) survey \citep{2017AJ....154...94M}, which are mostly thin and thick disk stars, and which include some ancient stars that likely formed in the MW progenitor and/or accreted halo stars. The VOYAGERS survey targets were selected to avoid those with [Fe/H] $\geq -0.8$ and [$\alpha$/Fe] values that overlap with the majority of thin and thick disk stars (grey points).
\textit{Right:} HR diagram for the VOYAGERS stars. For context, we overplot two 10\,Gyr MIST isochrones with $\mathrm{[Fe/H]}=-0.5$\,dex and $[\mathrm{Fe/H}]=-1.40$\,dex.}
\label{fig:met}
\end{figure*}

\begin{figure*}[!ht]
\centering{\includegraphics[width=0.9\textwidth]{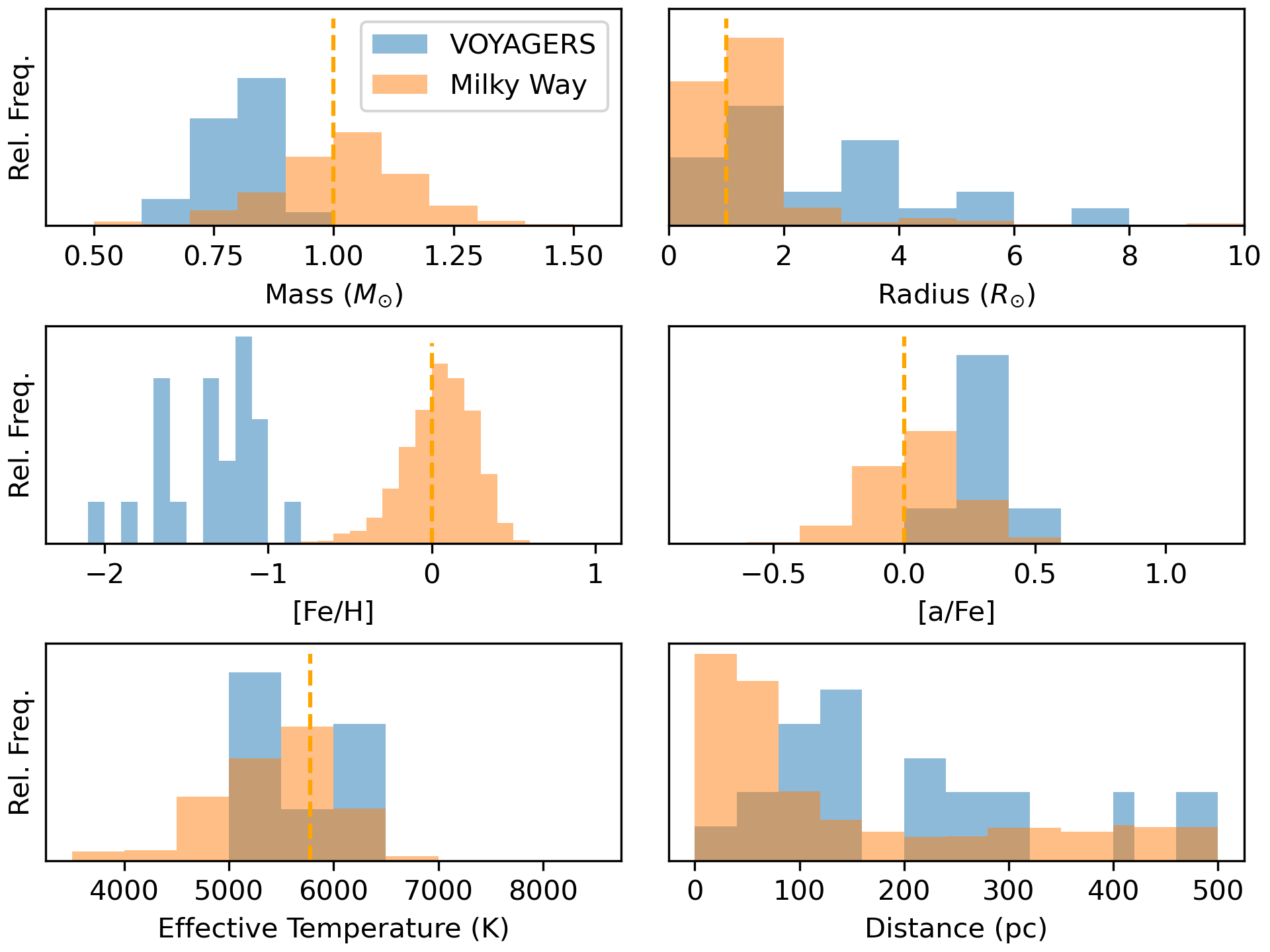}}
\caption{Stellar properties of VOYAGERS survey stars (blue bars) compared to properties of Milky Way planet-hosting stars (orange bars), as listed in the Hypatia Catalog of Planet Hosting Stars.  Solar properties are noted by orange dashed lines.  VOYAGERS survey stars are metal poor and alpha element rich, with sub-solar mass and similar effective temperatures.  The distance to survey stars is typical compared to known planet hosting stars.}
\label{fig:stellar}
\end{figure*}

The stellar properties of our targets are summarized in Table~\ref{deluxetable:stellar_properties} and in Figure~\ref{fig:met}.  Figure~\ref{fig:met}, (\textit{left}) compares GES metallicities to those of thin and thick disk Milky Way stars and (\textit{right}) shows their location on an Hertzsprung-Russell (HR) diagram.  The target stars are typically around 10\,Gyr age, with masses ranging from $0.6-1.0\,M_{\odot}$.  More massive stars should have evolved into giants by the time they reach the age of GES.  The first 11 stars in the table are main sequence stars while the latter 11 stars are somewhat evolved with radii ranging from $2.5-7.2\,R_{\odot}$. The properties of GES stars are compared to those of Milky Way stars from the Hypatia Catalog of Planet Hosting Stars \citep{2014AJ....148...54H} in Figure~\ref{fig:stellar}.  GES stars included in our survey were selected as suitable for an RV survey, so these comparisons may not represent the properties of the larger GES population.
 
\begin{deluxetable*}{lccccccccc}
\label{deluxetable:stellar_properties}
\tablecaption{Stellar Parameters for VOYAGERS Targets\label{tab:parameters}}
\tablewidth{0pt}
\tablehead{
\colhead{} & \colhead{$T_{\mathrm{eff}}$} & \colhead{Mass} &
\colhead{Radius} & \colhead{$\log g$} & \colhead{[Fe/H]} & \colhead{[$\alpha$/Fe]} \\
\colhead{Name} & \colhead{(K)} & \colhead{(M$_\odot$)} & \colhead{(R$_\odot$)} & \colhead{(dex)}& \colhead{(dex)}& \colhead{(dex)}}
\startdata
\textit{Focused survey targets} \\
TIC 4624167 & $5445\pm130$ & $0.63\pm0.03$ & $0.61\pm\redit{0.03}$ & $4.67\pm0.23$ & $-1.7\pm0.09$ & $0.32\pm0.02$ \\
TIC 459799364 & $5571\pm133$ & $0.6\pm0.03$ & $0.81\pm\redit{0.03}$ & $4.4\pm0.22$ & $-1.37\pm0.07$ & $0.26\pm0.01$ \\
TIC 81103318 & $6106\pm148$ & $0.73\pm0.04$ & $0.96\pm\redit{0.04}$ & $4.34\pm0.22$ & $-1.62\pm0.08$ & $0.42\pm0.02$ \\
TIC 377249704 & $6270\pm151$ & $0.77\pm0.04$ & $1.01\pm\redit{0.04}$ & $4.32\pm0.22$ & $-1.68\pm0.08$ & $0.44\pm0.02$ \\
TIC 275153790 & $6440\pm155$ & $0.77\pm0.04$ & $1.01\pm\redit{0.04}$ & $4.31\pm0.22$ & $-2.09\pm0.11$ & $0.43\pm0.02$ \\
TIC 16041611 & $6276\pm151$ & $0.78\pm0.04$ & $1.04\pm\redit{0.04}$ & $4.29\pm0.21$ & $-1.64\pm0.08$ & $0.37\pm0.02$ \\
TIC 417581428 & $6157\pm148$ & $0.78\pm0.04$ & $1.06\pm\redit{0.05}$ & $4.28\pm0.21$ & $-1.4\pm0.07$ & $0.37\pm0.02$ \\
TIC 72052060 & $5821\pm140$ & $0.73\pm0.04$ & $0.85\pm\redit{0.04}$ & $4.44\pm0.22$ & $-1.02\pm0.05$ & $0.17\pm0.01$ \\
TIC 85335198 & $6354\pm153$ & $0.82\pm0.04$ & $1.22\pm\redit{0.05}$ & $4.18\pm0.21$ & $-1.53\pm0.08$ & $0.37\pm0.02$ \\
TIC 230241743 & $6177\pm149$ & $0.88\pm0.04$ & $1.27\pm\redit{0.06}$ & $4.18\pm0.21$ & $-0.88\pm0.05$ & $0.17\pm0.01$ \\
TIC 3806602 & $6224\pm150$ & $0.88\pm0.04$ & $1.42\pm\redit{0.06}$ & $4.07\pm0.2$ & $-1.12\pm0.06$ & $0.24\pm0.02$ \\
\hline
\textit{Additional survey targets}\\
TIC 88564693 & $5205\pm125$ & $0.86\pm0.04$ & $3.78\pm\redit{0.17}$ & $3.21\pm0.16$ & $-1.1\pm0.06$ & $0.19\pm0.01$ \\
TIC 450339861 & $5261\pm126$ & $0.79\pm0.04$ & $3.26\pm\redit{0.14}$ & $3.31\pm0.17$ & $-1.38\pm0.07$ & $0.33\pm0.02$ \\
TIC 164529844 & $5390\pm129$ & $0.71\pm0.04$ & $2.75\pm\redit{0.12}$ & $3.41\pm0.17$ & $-1.84\pm0.09$ & $0.32\pm0.02$ \\
TIC 94280898 & $5231\pm125$ & $0.87\pm0.05$ & $3.11\pm\redit{0.14}$ & $3.39\pm0.17$ & $-1.14\pm0.06$ & $0.35\pm0.02$ \\
TIC 284567525 & $5246\pm126$ & $0.81\pm0.04$ & $3.48\pm\redit{0.15}$ & $3.26\pm0.16$ & $-1.38\pm0.07$ & $0.36\pm0.02$ \\
TIC 357439577 & $5190\pm124$ & $0.88\pm0.04$ & $3.6\pm\redit{0.16}$ & $3.27\pm0.16$ & $-1.06\pm0.05$ & $0.29\pm0.02$ \\
TIC 18153046 & $5145\pm123$ & $0.82\pm0.05$ & $4.3\pm\redit{0.20}$ & $3.08\pm0.16$ & $-1.14\pm0.06$ & $0.25\pm0.01$ \\
TIC 165057488 & $5118\pm122$ & $0.88\pm0.05$ & $5.37\pm\redit{0.24}$ & $2.92\pm0.15$ & $-1.17\pm0.06$ & $0.28\pm0.02$ \\
TIC 232969674 & $5070\pm121$ & $0.87\pm0.05$ & $5.7\pm\redit{0.25}$ & $2.87\pm0.14$ & $-1.15\pm0.06$ & $0.34\pm0.02$ \\
TIC 157336760 & $5831\pm140$ & $0.95\pm0.05$ & $2.32\pm\redit{0.10}$ & $3.68\pm0.18$ & $-1.21\pm0.06$ & $0.21\pm0.01$ \\
TIC 20897763 & $5026\pm120$ & $0.87\pm0.05$ & $7.57\pm\redit{0.33}$ & $2.62\pm0.13$ & $-1.24\pm0.06$ & $0.29\pm0.02$ \\
\enddata

\tablecomments{Spectral properties calculated by \texttt{uberMS} using either TRES or HARPS-N spectra \redit{(2 targets where TRES spectra not available)}, isochrones, and available photometry.  \redit{Errors include floor errors from \citet{2022ApJ...927...31T} added in quadrature.}  The 10 targets selected for a focused survey moving forward as described in Section \ref{section:yield} are identified in the table.  The rows are organized with increasing stellar radii within groups.}
\end{deluxetable*}

\subsection{Selection of Survey Stars}
\label{sec:survey_selection}
From our screened list of 41 stars, we selected 22 stars to include in the survey.  The number of targets was selected following the estimated yield calculations described in Section \ref{section:yield}.  The 22 stars included 11 stars with $R<2.5\,\mathrm{R_\odot}$ (these have the lowest RV scatter due to processes like granulation and pulsations), and 11 more evolved stars ($R <7.5\,\mathrm{R_\odot}$) to increase our ability to detect less common, but higher minimum mass planets during the survey.  We divide the stars into the 11 main sequence stars and the 11 more evolved stars. We have avoided the selection of GES stars with $\mathrm{[Fe/H]} \geq -0.8$ that could overlap the metallicity of the thick disk.

\section{Estimated Survey Yield}
\label{section:yield}

A key goal of our survey is not just to detect planets that were born outside the Milky Way, but also to be a first probe of their population and to see whether there are significant differences between GES planets and Milky Way planets. Therefore, to design the survey, we need to estimate how many planets we would detect if the populations were identical between GES and Milky Way, and set the number of targets and observations such that even if the survey detects no planets, we still have a significant result.  We note that the RV analysis gives mass multiplied by $\sin (inclination)$ or the minimum mass for exoplanets.  We account for random inclinations in our yield models

We therefore conducted the following analysis to predict the yield of planets from our survey. 
Specifically, we took the following steps to estimate our survey yields:
\vspace{-2.0mm}
\begin{itemize}
\item{We generated a random distribution of planet radii and periods for both G- and K-type stars for 500 systems each. These distributions are generated from the Kepler mission planet occurrence rates, as calculated by \citet{2023AJ....166..122D}. }  
\vspace{-2.0mm}
\item{For each search, the stellar mass from the VOYAGERS target stars is combined with a random planet radius and period.  Planet minimum mass is calculated using the non-parametric model from \citet{2018ApJ...869....5N}.  The minimum mass is reduced to 50 M$_{\oplus}$ for all more massive planets as a conservative estimate because we did not expect to find more massive gas giants orbiting these metal-poor stars.  We note that the maximum mass of planets in GES is not observationally constrained, so our choice of 50 M$_{\oplus}$ is just a reasonable placeholder estimate. Ultimately, this value will be constrained by future analysis of GES planet occurrence.  We set minimum period for each search to the period at which the ratio of semi-major axis / stellar radius is 4 to account for the fact that some of our stars are evolved with up to 7.5 R$_{\odot}$.}
\vspace{-2.0mm}
\item{An RV semi-amplitude is calculated based on the stellar mass, planet mass and period, assuming a circular orbit.  The impact of orbital inclinations with respect to Earth ($i$) is simulated by multiplying the observed semi-amplitude ($K_{\mathrm{obs}}$) by $\sin(i)$ where $i$ is generated by taking the inverse cosine of a random number drawn uniformly between 0 and 1.}
\vspace{-2.0mm}
\item{\redit{As shown in Equation \ref{eq:1}, we} compare $K_{\mathrm{obs}}$ to the RV$_\mathrm{{rms}}$ sensitivity of a target star randomly drawn from main sequence or evolved survey stars as appropriate.}  

\begin{equation} \label{eq:1}
        10 < \frac{K_{\mathrm{obs}}\mathcal{*}\sin(i)} {(\redit{RV}_{\mathrm{rms}} / \sqrt{N})}
\end{equation}

The number of RV values for each target is N.  We use a target signal to noise ratio (S/R) of 10 based on the S/R needed to constrain false alarm probability (FAP) to below 1\% using injection/recovery experiments with actual survey data. 
We use the actual achieved RV$_{\mathrm{rms}}$ from our observations for a randomly selected star.  This assumes that the variability is random noise, but RV$_{\mathrm{rms}}$ may include Keplerian signals, so these assumptions may underestimate the S/R.  

\vspace{-2.0mm}
\item{We included the potential impact of complex RV signals on discovery of compact multi-planet systems \citep{2018ApJ...867L...3B} by doubling the RV$_{rms}$ for a random 15\% of planets with $<$ 10\,M$_{\oplus}$.  If the planned S/R is achieved, we estimate two planet detections for such systems.}
\vspace{-2.0mm}
\item {We repeated these simulated surveys 10,000 times to determine the range of possible outcomes.}
\end{itemize}

\begin{figure}[!ht]
\centering{\includegraphics[width=0.5\textwidth]{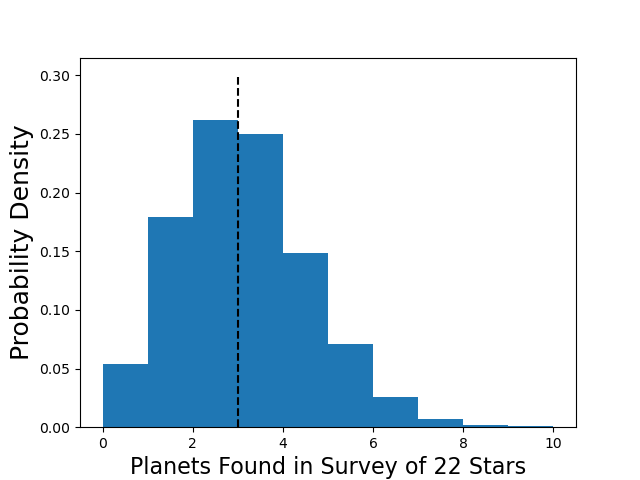}}
\caption{Expected survey yield of $> 5\,M_{\oplus}\sin{i}$ planets for 22 stars based on 10,000 simulated searches with 115 observations per star, 10 S/R, and random RV$_\mathrm{rms}$ precision drawn from actual observation results.  The dashed vertical line indicates our median yield of of three planets.}
\label{fig:yield}
\end{figure}

Given the assumptions above, which include occurrence rates similar to those measured in the Milky Way, the median expected yield of a survey of 11 main sequence stars and 11 evolved stars is estimated to be three $> 5\,M_{\oplus}\sin{i}$ planets with 95\% confidence of detecting at least one planet if we collect 115 observations on each target.  The distribution of yield results is shown in Figure~\ref{fig:yield}.  The total number of observations (2530) this survey would require based on awarded observing time from recent semesters results in a survey period of approximately 10 years.

\begin{figure}[!ht]
\centering{\includegraphics[width=0.5\textwidth]{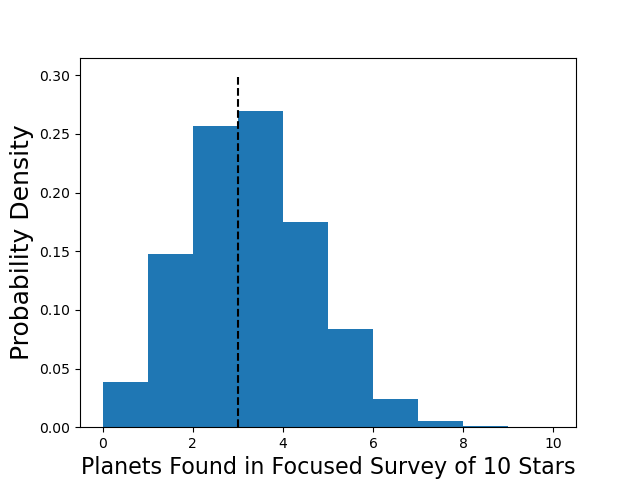}}
\caption{Expected survey yield of $> 5\,M_{\oplus}\sin{i}$ planets for 10 stars based on 10,000 simulated searches with 160 observations per star, 10 S/R, and random RV$_\mathrm{rms}$ precision drawn from actual observation results.  The dashed vertical line indicates our median yield of of three planets.}
\label{fig:yield_ms}
\end{figure}

\begin{figure}[!ht]
\centering{\includegraphics[width=0.5\textwidth]{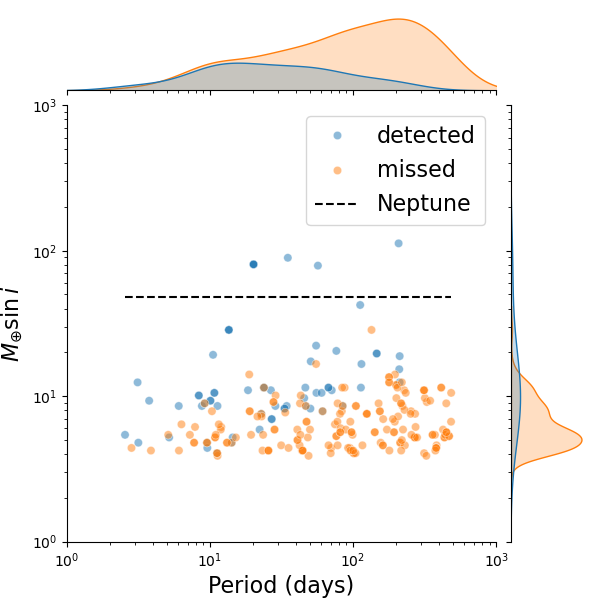}}
\caption{Period - mass diagram showing distribution of 250 simulated planets detected and missed during 10 main \redit{sequence} star yield analysis.  Blue circles indicate planets meeting the detection criteria while orange circles did not meet the criteria.  The histograms show that survey detected planets peaked at shorter periods (\textit{top}) and higher masses (\textit{right}) compared to missed planets, as expected.}
\label{fig:period_mass}
\end{figure}

We note that we could improve our sensitivity to lower-mass exoplanets and potentially expedite earlier detections if we focus future observations on our main sequence targets.  We repeated yield calculations for 10 main sequence stars.  A candidate, TIC 275153790, was not included in this analysis because it has the longest exposure time at 2250 s and still has a relatively high RV$_\mathrm{rms}$ of 4.1 $\,m\,s^{-1}$ compared to a mean of 2.1 $\,m\,s^{-1}$ for the other main sequence targets.  No promising periodic signals are present after 36 observations of TIC 275153790.  
Our simulations indicated that 160 observations of 10 stars would also yield three  $> 5\,M_{\oplus}\sin{i}$ planets with 95\% confidence of detecting at least one planet (see Figure~\ref{fig:yield_ms}), reducing the total observations to 1600 while meeting the same science goals.  We therefore plan to proceed with a focused survey of 10 targets, with the remaining 12 targets retained to observe during sub-optimal conditions.
An example of the period - mass distribution for simulated planets detected during the focused survey yield analysis is shown in Figure \ref{fig:period_mass} for 25 simulations.  The mass of Neptune is shown as a reference.  No maximum planet mass was used for this figure.  The histogram of detected planets on the top peaks at shorter periods than missed planets, while the histogram on the right shows a wide range of masses for detected planets, while missed planets peak at lower mass.  The observed semi-amplitudes of the detected planets ranged from 1.2 to 34.8 $\,m\,s^{-1}$.

\section{Precision Radial Velocity Observations}
\label{section:rv}

We are collecting RVs using three precision radial velocity spectrographs: NEID \citep{2016SPIE.9908E..7HS}, HARPS-N \cite[the High Accuracy Radial velocity Planet Searcher--North;][]{2012SPIE.8446E..1VC}, and CARMENES \cite[Calar Alto high-Resolution search for M dwarfs with Exo-Earths with Near-infrared and optical Échelle Spectrographs;][]{2016SPIE.9908E..12Q}.  
Here, we describe each spectrograph and the observations in more detail. 

\subsection{Radial Velocity Instruments}
\subsubsection{NEID}
\label{subsubsection:NEID}
\redit{NEID} is a high resolution echelle spectrograph on the 3.5-meter WIYN telescope on Kitt Peak National Observatory in Arizona. 
The resolution of NEID using the high resolution (HR) fiber is approximately 110,000, while a larger diameter high efficiency (HE) fiber has a resolution of 72,000. Both fibers cover a wavelength range of 380--930\,nm.  The instrument is maintained in an ultrastable environment with $< 1\,\mathrm{mK}$ temperature variance at (${\sim}180\,K$) and pressure $< 10^{-6}\,\mathrm{Torr}$ \redit{These parameters are described in \citet{2016SPIE.9908E..7HS, 2019JATIS...5a5003R}}.
Initial NEID exposure times were based on estimates using the NEID Exposure Time Calculator \redit{(ETC)} to target $\leq 2.0\,m\,s^{-1}$ \redit{photon noise error} (S/R${\sim}100$).  \redit{The NEID ETC exposure times assume solar metallicity.  We did find} that longer exposure times were required to achieve this photon noise error, likely due to the low metallicity and shallower spectral lines for our target stars.  During the 2022B observing semester, observations using the HE fiber were found to provide better RV sensitivity than NEID observations using the smaller diameter HR fiber.  All observations since then have utilized the HE fiber.  NEID exposure times with the HE fiber for main sequence targets range from 600--2500 seconds and currently average $3.1\,\mathrm{m\,s}^{-1}$ RV$_{\mathrm{error}}$ (using the \texttt{SERVAL} pipeline discussed below). NEID exposures for slightly evolved targets range from 600--1200 seconds and average $4.1\,\mathrm{m\,s}^{-1}$ RV$_{\mathrm{error}}$.
As discussed in Section~\ref{sec:survey_selection}, evolved stars exhibit higher RV jitter; therefore, we employ shorter exposures and tolerate higher RV$_{\mathrm{error}}$.  RV and photon noise error are calculated utilizing the NEID instrument pipeline DRP 1.4.1 \redit{as well as the \texttt{SERVAL} code described in Section \ref{subsubsection:serval}}.  RVs are calculated with this pipeline using a cross-correlation function (CCF) method, \redit{comparing the spectra shifted by a range of RVs to masks derived from theoretical line lists to develop a CCF quantifying the similarity between the spectra and mask, then fitting a Gaussian to the peak of the CCF to determine the RV.}  VOYAGERS observations began in NEID Era 5 (October 18, 2022 to August 19, 2024) and continue in Era 6 (August 24, 2024 -- present).  NEID DRP 1.4.1 notes indicate that an RV offset is expected between eras 5 and 6 \redit{as described in \citet{2025arXiv250623704G}}.  Therefore, we analyze NEID data as three separate groups depending on fiber and era with separate RV offsets for each group (HR5, HE5, and HE6).

\subsubsection{HARPS-N}
HARPS-N is a high resolution echelle spectrograph on the 3.6-m Telescopio Nationale Galileo (TNG) telescope.  The resolution for HARPS-N is 115,000 covering wavelengths ranging from 383--690\,nm.  The instrument is maintained under high vacuum at ${\sim}300\,K$ with 1\,mK short term variability.  \redit{These instrument parameters are described in \citep{2012SPIE.8446E..1VC}}.  Like with NEID, the exposure times are increased relative to values suggested by the exposure time calculator due to the relatively shallow and few spectral features of our targets \redit{when compared to solar metallicity features assumed by the calculator}. Exposure times range from 800--2400\,s for main sequence stars and 600--1800\,s for evolved stars.  The instrument pipeline for HARPS-N uses CCFs to determine the RVs utilizing the HARPS-N Data Reduction Pipeline \citep{2021A&A...648A.103D}.  Data is accessed through the IA2 (Italian center for Astronomical Archives) facility and Data Analysis Center for Exoplanets (DACE, \citealt{2019ASPC..521..757B}) \footnote{\href{https://dace.unige.ch}
{https://dace.unige.ch}}.  We also download \redit{multi-order} spectra and use \texttt{SERVAL} to recalculate the RVs.

\subsubsection{CARMENES}
CARMENES is a high-resolution spectrograph on the 3.5-m telescope at the Calar Alto Observatory.  It has both a visible spectral arm (VIS) with a resolution of 94,600 ranging from 520--960\,nm and a near infrared (NIR) arm with a resolution of 80,000 ranging from 960--1710\,nm.  The VIS arm is maintained in a near vacuum at ${\sim}285$\,K with 10\,mK short term stability. \redit{Parameters are described in \citep{2016SPIE.9908E..12Q}}.  Our CARMENES RVs are calculated using the VIS observations as they have been shown to exhibit lower RV errors; however, NIR results are also available for each observation.  We are targeting the CARMENES observations to the lowest $T_{\mathrm{eff}}$ stars in the survey to take advantage of the spectrograph's high sensitivity at redder wavelengths.

We calculate CARMENES RVs using the \texttt{SERVAL} program as described in Section \ref{subsubsection:serval}.  Initial CARMENES observations had exposures resulting in average S/R of approximately 50--70; however, like the other spectrographs, we found that the CARMENES exposure time calculator underestimated the exposure time to achieve a given RV precision, so we increased the exposure times to target a higher S/R of up to 150 for main sequence targets and 100 for evolved targets, with exposures ranging between 750--2200 seconds.  

\subsubsection{\redit{Archival Observations}}
\redit{There are several targets with publicly available archival observations using HIRES, HARPS, and ESPRESSO instruments.  We include this data in our analysis of potential exoplanet signals.} 

HIRES is a high-resolution spectrograph on the 10-m Keck telescope at the Mauna Kea Observatory in Hawaii.  It has a resolution of 84,000 and a spectral range from 300--1100\,nm.  The instrument is maintained at ${\sim}143\,K$ \redit{as described in \citet{1994SPIE.2198..362V}}.  We use archival observations with S/R of 25 or greater found in \citet{2025A&A...702A..68T} for \redit{TIC 4624167 (18 observations over 8071 days ), TIC 459799364 (13 observations over 7474 days, and TIC 72052060 (23 observations over 7310 days) when searching for periodic signals.}

\redit{HARPS is a high-resolution spectrograph on the European Southern Observatory (ESO) 3.6-m telescope at the ESO La Silla observatory.  It has a resolution of 90,000 and a spectral range from 380\,nm to 680\,nm.  The instrument is described in \citet{2000SPIE.4008..582P}.  We use archival observations available from \citet{2024A&A...683A.125P} for TIC 377249704 (6 observations binned over 2 nights) and TIC 72052060 (7 observations over 3958 days).}

\redit{ESPRESSO is a high-resolution spectrograph on the ESO Very-Large Telescope (VLT). It has a resolution of a spectral resolving power of 140,000 or 190,000 over the 378.2 to 788.7 nm wavelength range.  The instrument is described in \citet{Pepe_2021}.  We use archival observations available for TIC 72052060 that were available from the ESO Archive using DACE (24 observations over 128 nights).}

\redit{
\subsubsection{\texttt{SERVAL}}
\label{subsubsection:serval}
We use the \texttt{SERVAL} code \citep{2018A&A...609A..12Z} to calculate RVs in addition to the instrument pipelines discussed above. \texttt{SERVAL} calculates RVs based on an iterative approach in which a co-added spectrum is produced from the individual observations, shifting each spectrum according to an RV estimate. The RV shift for each observation is then calculated relative to this co-added spectral \redit{template} using a least-squares fit. 
We use a \redit{modified} version, \texttt{NEIDSERVAL}, as described in \citet{2022ApJ...931L..15S}, on NEID data. The \texttt{SERVAL} approach was used to calculate RVs for all three spectrographs.  \redit{We use the following activity indicators from \texttt{SERVAL} as described in \citet{2018A&A...609A..12Z} to evaluate whether RV signals may correspond to stellar activity rather than Keplerian signals: Chromatic RV index (CRX), differential line width (dLW), H$\alpha$, Ca II IRT, and Na I D indices.}   
To mitigate telluric contamination, a custom telluric line mask has been developed using VOYAGERS data. The telluric filter, \texttt{mask\_0.01.txt}, was provided by the CARMENES team based on spectra from ten CARMENES observations for this survey. 
Observations with an average S/N below 75 for main sequence stars or below 25 for evolved stars were excluded from the analysis for all instruments. We track RVs calculated using both the CCF and \texttt{SERVAL} approaches when searching the data for periodic signals.}

\subsection{Status of Observations}

The VOYAGERS survey was initiated using the NEID spectrograph on the WIYN 3.5-m telescope in semesters 2022B and 2023A \redit{(beginning November 12, 2022)} using University of Wisconsin--Madison institutional time.  In addition, NNEXPLORE proposals using WIYN/NEID were added /redit{starting September 4, 2023)} in 2023B to increase the number of observations. Since then, we have also been awarded time and have started observations on \redit{August 1, 2023} with CARMENES and \redit{on August 9, 2023} with HARPS-N starting in the 2023B semester.  Our final survey analysis will incorporate RV observations from the three spectrographs.  As of June 10, 2025, 778 observations have been completed for the VOYAGERS survey, including 519 NEID, 166 HARPS-N, and 93 CARMENES observations on 22 targets.  We have completed 385 observations on the 10 main sequence stars identified for the focused survey, including 256 NEID, 86 HARPS-N, and 43 CARMENES observations.   

\section{Initial Exoplanet Search}
\label{section:search}

As survey observations progress, we continuously monitor the incoming data to confirm the quality is high and adjust observing strategy as needed. To perform quick analyses on the fly, we use the DACE online interface, managed by the University of Geneva \citep{2019ASPC..521..757B}. DACE allows nearly real-time visualization of the radial velocity observations, periodogram searches for radial velocity planet candidates, and includes tools to perform quick fits of the relative RV data to obtain an orbital solution.  We upload the \texttt{SERVAL} output for each instrument and star into DACE, including the RV measurements, their uncertainties, and a series of activity indicators including line shape diagnostics \redit{(such as FWHM or Bisector Span from CCF instrument pipelines)} and the strengths of magnetically sensitive lines.  We review these stellar activity index values to search for correlations with RV measurements and discriminate stellar signals from true planetary RV shifts.

The procedure we use to quickly vet signals is as follows: 
\begin{enumerate}
    \item We import the time series for each instrument into DACE, and fit for an offset value for each instrument. Because the NEID data have different observation epochs, we include separate offsets calculated for each (corresponding to the HR5, HE5, and HE6 datasets).
    \item We calculate the Generalized Lomb Scargle periodogram \citep{Lomb1976Ap&SS, Scargle1982ApJ, Zechmeister2009A&A} of the combined datasets (with the offsets included). We search periods from 0.1 to 10,000 days.  We identify periods with power exceeding analytically calculated false alarm probabilities (FAP) levels of 10\%, 1\% and 0.1\%, indicating detections of increasingly significant periodicity following \citet{2008MNRAS.385.1279B}. 
    \item Once we have identified a potential Doppler periodicity, we then use DACE to fit selected periods with one or more Keplerian models and/or a stellar jitter term. In cases where multiple potentially significant signals are detected, we can model multiple Keplerian signals simultaneously.
    \item \redit{We similarly search timeseries for the activity indicators outlined in Section \ref{subsubsection:serval} for significant periodic signals and we compare any periods close to RV periods that may have been related to stellar activity.  We are able to follow up by investigating correlations between RV and activity indicator signals.}
\end{enumerate} 


\section{Discussion}
\label{section:discussion}

\subsection{Sensitivity to Low Mass Exoplanets}
\label{subsection:lowmass}

\begin{figure*}[!t]
\centering{\includegraphics[width=0.8\textwidth]{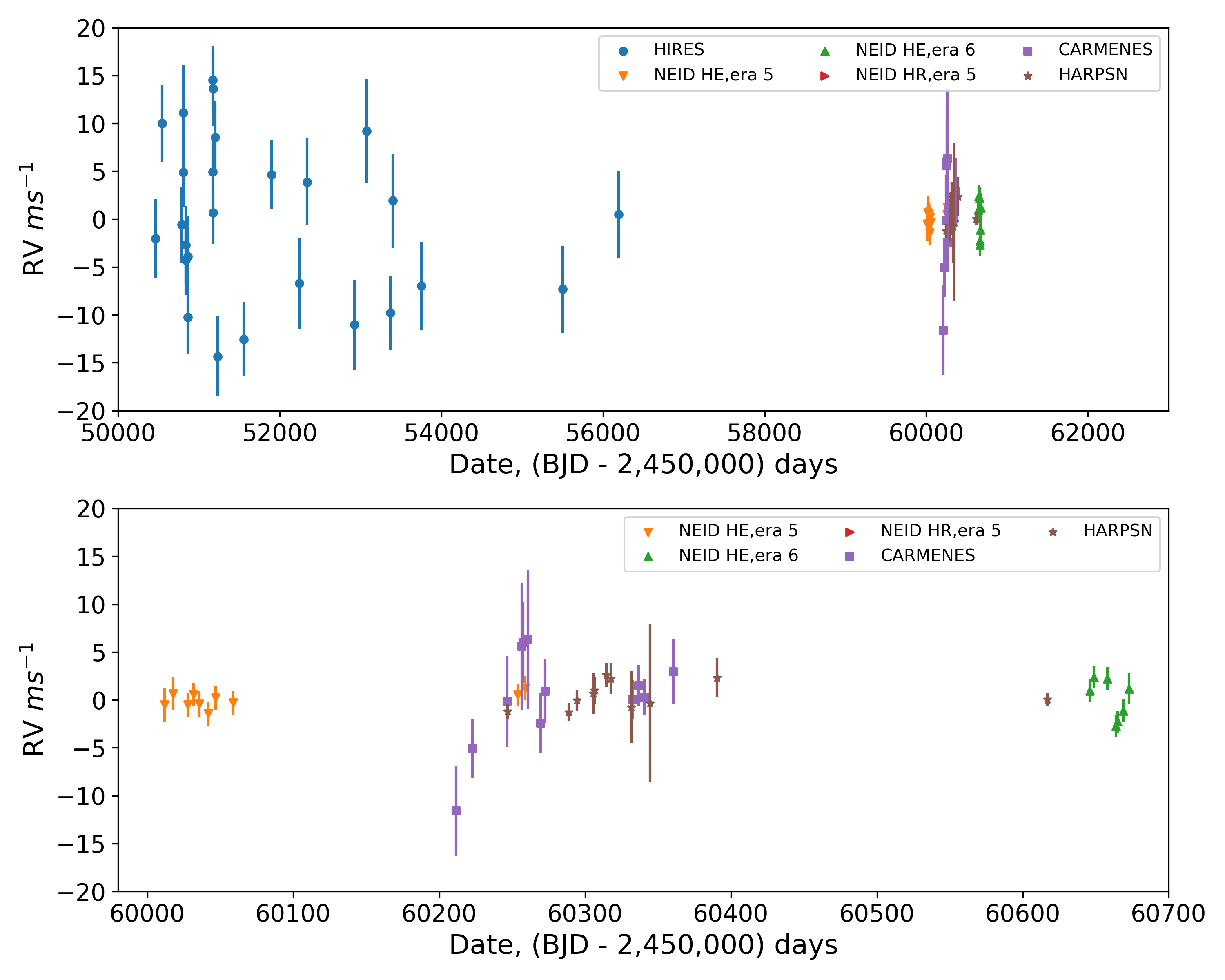}}
\caption{(\textit{top}) Example time series of TIC 4624167 with 50 RV observations from NEID, HARPS-N, CARMENES, as well as \redit{18 archived HIRES observations}.  (\textit{bottom}) Focus on recent survey observations.  No significant periodic signals were found.}
\label{fig:timeseries}
\end{figure*}

We designed the VOYAGERS survey with the intention of achieving sensitivity to sub-Neptune mass exoplanets orbiting GES members. In order to assess our ability to achieve this science goal, we conducted an injection/recovery analysis using time series survey data given our achieved RV precision and sampling of observations.  
In Figure~\ref{fig:timeseries}, we show the time series RV observations from combination of NEID, HARPS-N, and CARMENES for one of the VOYAGERS targets, TIC 4624167.  Also shown are archival RVs from HIRES observations.  No periodic signals have yet been identified in this time series data at the 10\% FAP level.  

\begin{figure*}[!t]
\centering{\includegraphics[width=0.85\textwidth]{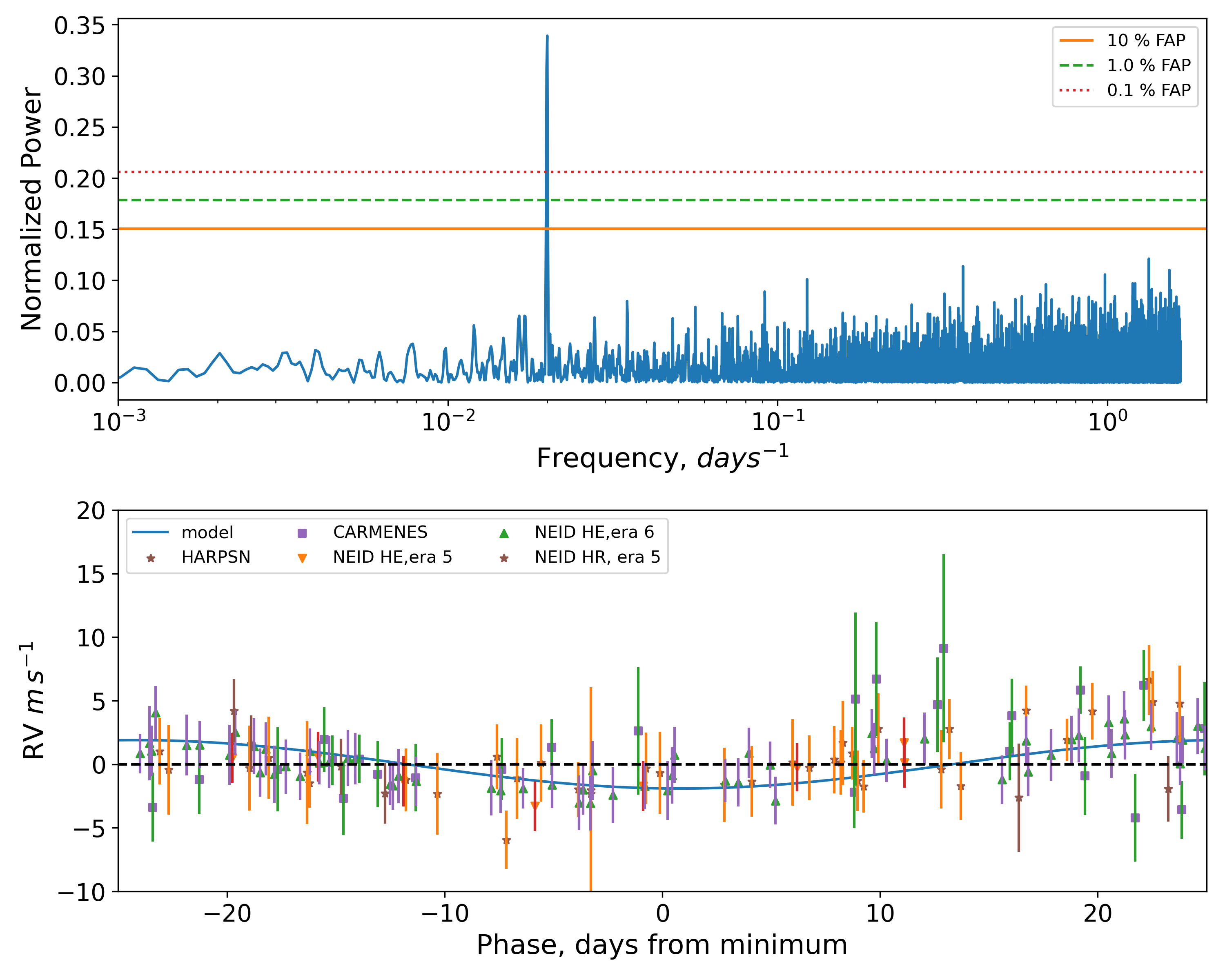}}
\caption{A 4.0\,m\,s$^{-1}$, 10 day period signal was injected into 50 actual observations and 110 simulated observations of TIC 4624167 from NEID, HARPS-N and CARMENES.  (\textit{top}) Periodogram of simulated observations showing detection of injected exoplanet at a 10$-$day period exceeding 1 \% FAP. (\textit{bottom}) Time series with injected exoplanet folded at 10 day period.  The recovered semi-amplitude is 4.5\,m\,s$^{-1}$.}
\label{fig:injected_1p}
\end{figure*}

\redit{To confirm that a signal could be recovered, we injected a 10-day periodic signal of 4.0\,m\,s$^{-1}$ (simulating an exoplanet with a minimum mass of 10.2\,M$_{\oplus}$) into 50 actual observations and 110 simulated observations of RV data for TIC 4624167.}  \redit{We simulated additional observations using the median RV and RV median absolute deviation to estimate the RV distribution.  Random RVs are generated modeled as a normal distribution at random time points between the most recent observation and 4 years afterwards.  We use the median RV error from actual observations to estimate the RV error for simulated observations.  On targets with CARMENES observations, we add 25\% of simulated data based on CARMENES observations.  We similarly add 25\% of simulated data based on HARPS-N observations, which are available for all targets.  The remaining 50\% to 75\% of data are simulated based on NEID HE fiber Era 6 observations so that we have a total of 160 observations.}  
Archival HIRES observations were excluded for the injection/recovery analysis, which focuses on the current survey.  The top plot of Figure~\ref{fig:injected_1p} is a periodogram of the data with the injected signal, showing the normalized power signal.  \redit{We searched periods from $0.6$\,d to $10000$\,d.}  To confirm the signal, the power of each period was analyzed to determine whether it exceeded 1\% FAP or log(FAP) of $-2.0$ for planet detection.

In our analysis, the most significant period identified was the injected \redit{10-day signal, with log(FAP) of $-4.6$.  The fit included offsets between instruments and a 5.5\,m\,s$^{-1}$ stellar jitter noise term in the DACE model.}  In the bottom plot of Figure~\ref{fig:injected_1p}, we show the phase-folded RV time series of the \redit{10-day} signal.  
\redit{The semi-amplitude for this signal is $4.5\,m\,s^{-1}$, implying a minimum exoplanet mass of $10.2\,M_{\oplus}$, which matches the injected value reasonably well.}  This example demonstrates that we should be able to detect sub-Neptune mass planets with orbital periods of tens of days.  We may also detect signals with similar or greater semi-amplitudes from larger mass planets with orbits up to hundreds of days as well as with various inclinations.  

\begin{figure*}[!t]
\centering{\includegraphics[width=0.85\textwidth]{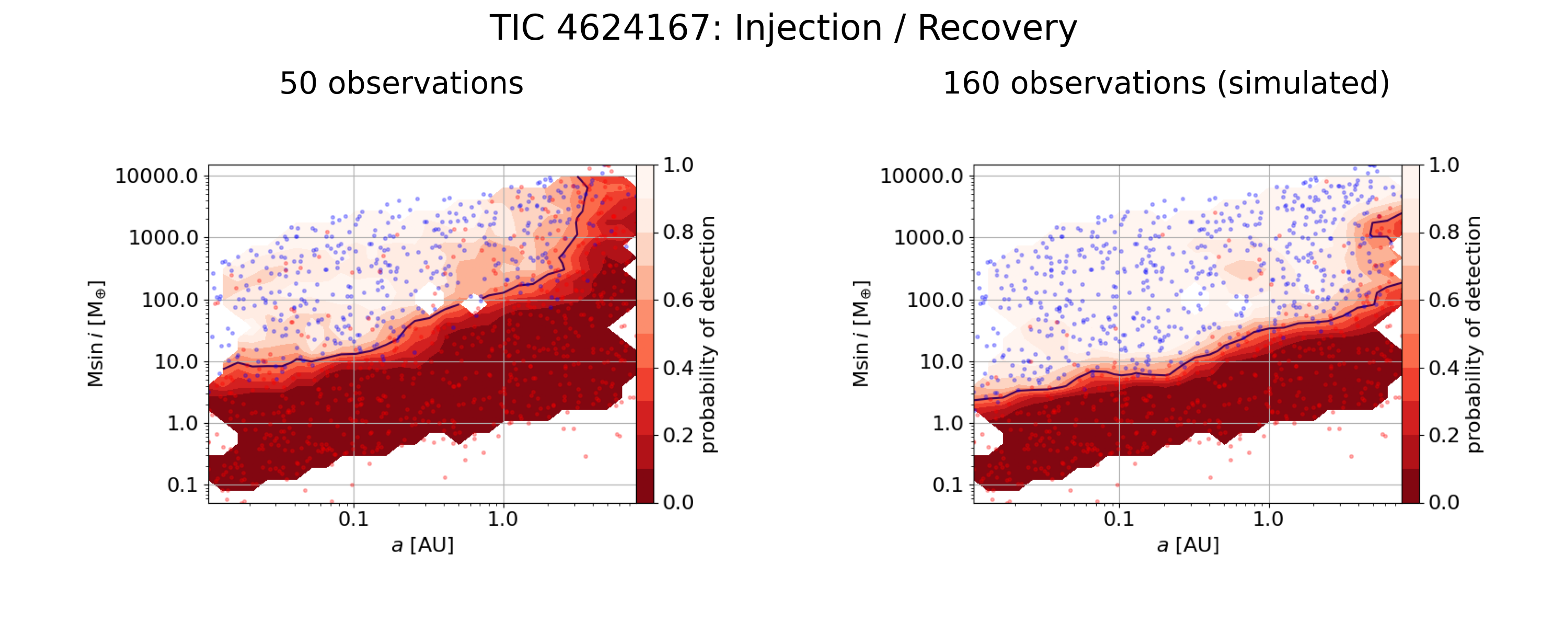}}
\caption{\redit{Completeness plots comparing fraction recovered from random signal injections into (\textit{left}) 50 actual and (\textit{right}) 160 actual and simulated observations of TIC 4624167.  Blue points represent detections while red points represent nondetections.  The solid black contour line separates the section of the graph with 50\% or better recovery from lower recovery rates indicated in darker red.  Note the improvement in the fraction recovered with 160 observations. }}
\label{fig:rvsearch_v016}
\end{figure*}

\redit{
\subsection{Injection / Recovery Test for Focus Survey}
To further explore sensitivity to low mass planets across the 10 targets included in the focused survey, we used the program \texttt{RVSEARCH} \citep{2021ApJS..255....8R} to inject 1000 random signals into actual survey data combined with simulated data totaling 160 observations for each target.  The 1000 random signals included randomly selected RV semi-amplitude ranging from $0.1$ to $1000\,m s^{-1}$, orbital periods from $0.5$ to $10,000$\,d, and eccentricities from $0.0$ to $1.0$.  The simulated data used the median RV and median absolute deviation of RVs as estimates of RV distribution.  We use the median of the observation RV error estimates to estimate RV errors for simulated data.  Random RVs are generated based on this distribution at random time points between the most recent observation and 4 years afterwards.  Figure \ref{fig:rvsearch_v016} compares the recovery percentage of 1000 random signals with the current 50 observations for TIC 4624167 to the percentage recovered after 160 observations.  As the number of observations increases to 160, exoplanets with lower mass and longer periods have 50\% or better recovery.  For the 160 observations plot, a $10\,M_{\oplus}$ mass exoplanet with a $0.1\,AU$ semi-major axis orbit would be recovered 76\% of the time, while a $50\,M_{\oplus}$ exoplanet with a $1.0\,AU$ orbit would be recovered 75\% of the time.  The remaining 9 targets in the focused study were  analyzed similarly.  The results are summarized in Table \ref{tab:percent_recovery}.  The recovery percentage for the 12 targets not included in the focused study ranged from 6 to 81\% for a $0.1\,AU$ semi-major axis orbit and 9 to 88\% for a $50\,M_{\oplus}$ exoplanet with a $1.0\,AU$.} 

\begin{deluxetable*}{lcc}
\label{tab:percent_recovery}
\tablecaption{\redit{Focused Survey  - Percent Recovery from Random Injections}}
\tablewidth{0pt}
\tablehead{
\colhead{\redit{Target}} & \colhead{\redit{$10\,M_{\oplus}$, $0.1\,AU$}} & \colhead{\redit{$50\,M_{\oplus}$, $1.0\,AU$}}
}
\startdata
\redit{TIC 4624167} & \redit{$76$} & \redit{$75$} \\
\redit{TIC 459799364} & \redit{$94$} & \redit{$89$}  \\
\redit{TIC 81103318} & \redit{$70$} & \redit{$63$}  \\
\redit{TIC 377249704} & \redit{$94$} & \redit{$93$} \\
\redit{TIC 16041611} & \redit{$85$} & \redit{$86$}  \\
\redit{TIC 417581428} & \redit{$97$} & \redit{$92$}  \\
\redit{TIC 72052060} & \redit{$78$} & \redit{$98$}  \\
\redit{TIC 85335198} & \redit{$88$} & \redit{$91$}  \\
\redit{TIC 230241743} & \redit{$96$} & \redit{$91$}  \\
\redit{TIC 3806602} & \redit{$85$} & \redit{$82$}  \\
\enddata
\tablecomments{\redit{The 10 targets selected for the focused survey all have greater than 50\% recovery for the examples analyzed.}}
\end{deluxetable*}

\subsection{Prospects to Quantify GES Elemental Abundances}
\label{subsection:GES Elements}

\redit{
In the process of collecting our survey data, we will observe the spectra of our GES target stars many times ($\sim 10^2 $ observations each) at high signal-to-noise ratios (S/N $\sim 10^2$ per pixel). An opportunity provided by the large number of high signal-to-noise observations is that we could co-add the observations together to create an extremely high signal-to-noise ratio spectrum of each star (with S/N $\sim 10^3$ per pixel). These extremely high S/N spectra would be excellent for measuring the detailed elemental abundances of stars in the GES. A similar strategy has been used before by numerous groups \citep{Mortier2013A&A, Bedell2018ApJ} for precise abundances of stars with metallicities closer to the Sun. For our survey, though, which targets stars with very different abundances than the Sun, it could be possible to measure shallow lines of elements that have too low abundances to detect in single exposures.  \redit{A discussion of how detailed elemental abundances can be used to explore GES star formation history can be found in \citet{2024A&A...691A.333E}, where [Fe/Mg], [Ba/Mg], [Eu/Mg], and [Eu/Ba], as a function of [Fe/H] are derived from the Stellar Abundances for Galactic Archaeology Database (SAGA, \citealt{2008PASJ...60.1159S}).  The high precision abundances for our targets will help determine how these low metallicity, potentially planet hosting stars formed.}  Ultimately, a by-product of our survey could be one of the most detailed chemical inventories of stars born outside the Milky Way.} 

\subsection{Challenges of Doppler surveys of low-metallicity stars}
Studying the population of exoplanet\redit{s} around low metallicity stars is \redit{a }potentially powerful way to constrain formation processes, how much solid material is needed to form planetary cores, and ultimately learn how early in the Universe's history planets and life could emerge. However, the low metallicity of the targets makes searching for planets via radial velocities challenging. In the course of this work, we \redit{identified} several impediments to designing an effective survey for planets around GES members that would also apply to other potential surveys of low-metallicity stars. 

\begin{figure*}[!t]
\centering{\includegraphics[width=1\textwidth]{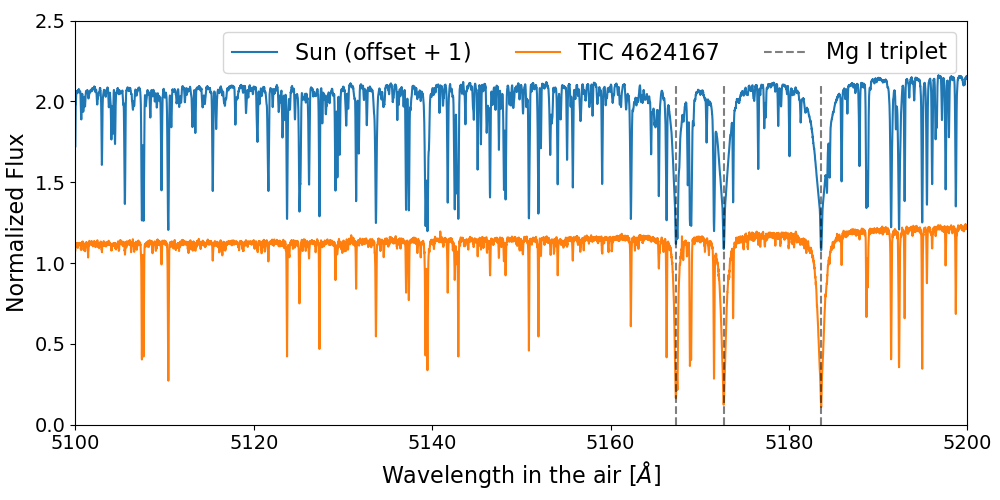}}
\caption{A narrow spectral window from a HARPS-N observation of TIC 4624167 is compared to a HARPS-N solar spectral window from three years of co-added spectral data.  The Mg I triplet is indicated by dashed vertical lines.}
\label{fig:spectra}
\end{figure*}

One key challenge is that low-metallicity stars have weaker spectral lines. Figure \ref{fig:spectra} compares a narrow spectral window for a single HARPS-N spectr\redit{um} of TIC 4624267 to a HARPS-N solar spectr\redit{um} from 3 years of observations \citep{2021A&A...648A.103D}. Many of the deep lines present in the solar spectr\redit{um} are either much shallower or absent in spectra of stars from the GES. Because the signal-to-noise limit of radial velocity measurements depends on the number and depth of absorption lines within a spectrum \citep{Connes1985Ap&SS, Bouchy2001A&A}, the photon-limited radial velocity precision we achieve is worse than an equivalently bright solar-metallicity star. This restricts our sample to relatively bright stars (typically $V \lesssim 10$) in order to keep exposure times reasonably short.  

The greater challenge of measuring precise RVs for low-metallicity stars compounds with the fact that GES stars (and extremely low-metallicity stars in general) are relatively rare in the Solar neighborhood. We were able to take advantage of a handful of main sequence stars with $V < 10$ currently known to be GES members, but we were unable to practically expand our survey to larger samples given the long exposure times needed to achieve sufficient signal-to-noise ratios with our current instruments. 


\section{Conclusions}
\label{section:conclusions}

In this work, we describe the VOYAGERS survey, which is a radial velocity survey designed to identify exoplanets orbiting 22 GES stars that were born in the Enceladus dwarf galaxy and are currently close enough to the Sun for PRV observations.  We designed the survey to be sensitive to sub-Neptune mass exoplanets around these metal-poor stars, assuming similar occurrence rates to Milky Way stars.   
We have formed a global collaboration, incorporating observations from NEID, HARPS-N, and CARMENES, to obtain high-precision RV measurements.
Our goal is to obtain 160 observations for each GES target.  We have completed 778 observations.  We will focus future observations on 10 main sequence targets to expedite results, while continuing to observe the remaining targets during sub-optimal conditions.  
Further, the survey is designed such that if we detect no planets, we will be able to determine with confidence that occurrence rates for Neptune-mass exoplanets are significantly lower for GES targets than occurrence rates for stars born in the Milky Way. 
If we discover one or more planets orbiting these ancient, low-metallicity stars, the survey results will extend our understanding of when and where planets and potentially life can evolve in the Universe.  

\vspace{1cm}

\textit{Acknowledgments:}
The authors thank the referee for their suggestions to improve the paper.  Data presented in this paper are based on observations obtained at the Fred Lawrence Whipple Observatory of SAO.
Based in part on observations at Kitt Peak National Observatory, NSF's National Optical-Infrared Astronomy Research Laboratory (NOIRLab Prop. 2022B-108242, 2023A-744382, 2023B-201966, 2024A-586095, 2024B-784368, 2025A-526298), which is operated by the Association of Universities for Research in Astronomy (AURA) under a cooperative agreement with the National Science Foundation and from WIYN telescope time allocated to NN-EXPLORE Prop.~2023B-788144, 2024A-272272, 2024B-670839, 2025A-234602 through the scientific partnership of the National Aeronautics and Space Administration, the National Science Foundation, and the NOIRLab.  The WIYN Observatory is a joint facility of the NSF’s National Optical-Infrared Astronomy Research Laboratory, Indiana University, the University of Wisconsin-Madison, Pennsylvania State University, and Princeton University.  This work is based on observations made with the Italian Telescopio Nazionale Galileo (TNG) operated on the island of La Palma by the Fundación Galileo Galilei of the INAF at the Spanish Observatorio del Roque de los Muchachos of the Instituto de Astrofisica de Canarias (GTO programme). The HARPS-N project was funded by the Prodex Program of the Swiss Space Office (SSO), the Harvard-University Origin of Life Initiative (HUOLI), the Scottish Universities Physics Alliance (SUPA), the University of Geneva, the Smithsonian Astrophysical Observatory (SAO), the Italian National Astrophysical Institute (INAF), the University of St. Andrews, Queen’s University Belfast and the University of Edinburgh.
Based in part on observations at CARMENES is an instrument at the Centro Astronómico Hispano en Andalucía (CAHA) at Calar Alto (Almería, Spain), operated jointly by the Junta de Andalucía and the Instituto de Astrofísica de Andalucía (CSIC). CARMENES was funded by the Max-Planck-Gesellschaft (MPG), the Consejo Superior de Investigaciones Científicas (CSIC), the Ministerio de Economía y Competitividad (MINECO) and the European Regional Development Fund (ERDF) through projects FICTS-2011-02, ICTS-2017-07-CAHA-4, and CAHA16-CE-3978, and the members of the CARMENES Consortium (Max-Planck-Institut für Astronomie, Instituto de Astrofísica de Andalucía, Landessternwarte Königstuhl, Institut de Ciències de l’Espai, Institut für Astrophysik Göttingen, Universidad Complutense de Madrid, Thüringer Landessternwarte Tautenburg, Instituto de Astrofísica de Canarias, Hamburger Sternwarte, Centro de Astrobiología and Centro Astronómico Hispano-Alemán), with additional contributions by the MINECO, the Deutsche Forschungsgemeinschaft through the Major Research Instrumentation Programme and Research Unit FOR2544 “Blue Planets around Red Stars”, the Klaus Tschira Stiftung, the states of Baden-Württemberg and Niedersachsen, and by the Junta de Andalucía.  The authors wish to thank Evangelos Nagel for assistance with a telluric mask and correcting CARMENES spectra for telluric contamination using molecfit. This publication makes use of The Data \& Analysis Center for Exoplanets (DACE), which is a facility based at the University of Geneva (CH) dedicated to extrasolar planets data visualization, exchange and analysis. DACE is a platform of the Swiss National Centre of Competence in Research (NCCR) PlanetS, federating the Swiss expertise in Exoplanet research. The DACE platform is available at https://dace.unige.ch.  This research has made use of the NASA Exoplanet Archive, which is operated by the California Institute of Technology, under contract with the National Aeronautics and Space Administration under the Exoplanet Exploration Program.  MP acknowledges financial support from the European Union – NextGenerationEU (PRIN MUR 2022 20229R43BH) and the ``Programma di Ricerca Fondamentale INAF 2023''.
AM acknowledges funding from a UKRI Future Leader Fellowship, grant number MR/X033244/1and a UK Science and Technology Facilities Council (STFC) small grant ST/Y002334/1.
HJD acknowledges support from the Spanish Research Agency of the Ministry of Science, Innovation and Universities (AEI-MICIU) under grant PID2023-149439NB-C41.
EP acknowledges financial support from the Agencia Estatal de Investigaci\'on of the Ministerio de Ciencia e Innovaci\'on MCIN/AEI/10.13039/501100011033 and the ERDF “A way of making Europe” through project PID2021-125627OB-C32, and from the Centre of Excellence “Severo Ochoa” award to the Instituto de Astrofisica de Canarias.
JCM acknowledges support through grant PID2021-125627OB-C31 funded by MCIU/AEI/10.13039/501100011033 and by “ERDF A way of making Europe”, and from the programme Unidad de Excelencia María de Maeztu CEX2020-001058-M, and from the Generalitat de Catalunya/CERCA programme, by the SGR 01526/2021.

\facilities{FLWO:1.5m (CfA Digital Speedometer, TRES), TNG (HARPS-N), WIYN (NEID), CALAR ALTO (CARMENES), Gaia}

\bibliography{sample631}{}
\bibliographystyle{aasjournal}



\end{document}